\documentclass[12pt]{article}
\usepackage{amssymb}

\usepackage{amsmath}

\input{tcilatex}

\begin{document}

\begin{center}
\bigskip

\bigskip

\bigskip

\bigskip

\bigskip

\bigskip \textbf{Relaxing the Differentiability Assumption in Taylor Theorem
and Optimization: Applications to Finance}

\bigskip 

Moawia Alghalith

UWI, St Augustine

malghalith@gmail.com

\bigskip

\bigskip
\end{center}

\textbf{Abstract}. We overcome a major obstacle in mathematical
optimization. In so doing, we provide smooth solutions to the HJB PDE
without assuming the smoothness of the value function$.$ We apply our method
to the portfolio model.

\bigskip


\bigskip \pagebreak 35Q92

\section{\protect\bigskip {\protect\large Introduction}}

A major obstacle in dynamic optimization is that the value function may not
be smooth (see, for example, Strulovici and Szydlowski (2015) for a
discussion). Actually, it is not expected to be smooth. A smooth solution to
the Hamilton-Jacobi-Bellman partial differential equation (HJB PDE) may not
exist. It is not surprising that a verification result exists only for a few
functional forms. In response, weak solutions such as viscosity solutions
were introduced (see, for example, Crandall and Lyon (1983), Hata and Sheu
(2012) and Gr\"{u}ne and Picarelli (2015), among many others).

In this paper, we overcome this obstacle in dynamic optimization. In doing
so, we present a simple method that relaxes the assumption of the
differentiability (smoothness) of the value function. That is, we generally
establish the existence and the uniqueness of a strong (smooth) solution
without the differentiability assumption.

We apply our method to three dominant models (the portfolio model, the
consumption-portfolio model, and the stochastic-factor model). However, the
extension to other areas is straightforward.\pagebreak

\bigskip

\section{\protect\Large Relaxing the differentiability assumption in Taylor
theorem}

\bigskip We introduce a pioneering approach that overcomes an obstacle in
mathematical sciences. In doing so, we introduce Taylor-like expansions even
if the original function is not differentiable. Needless to say, this result
is very useful in many applications, such as the areas of regression
analysis, optimization, integration, and partial differential equations PDEs.

Consider a continuous, bounded function $f\left( x\right) $ that is not
differentiable with respect to $x$ in a traditional sense; however, with a
simple transformation the function can become differentiable in some sense.
It can be expressed as $f\left( x+\beta \right) $, where $\beta $ is a shift
parameter with an initial value equal to zero. We define $i\equiv x+\beta $,
so that $f$ $\left( i\right) \equiv f\left( x+\beta \right) ,$ initially $%
f\left( i\right) =f\left( x\right) .$ We can show that $f$ is differentiable
(in some sense) w.r.t. $\beta $ (since every function can be shifted; this
can be easily seen graphically as a horizontal shift), and consequently $f$
is differentiable w.r.t. $i$ holding $x$ constant.

The idea is simple and intuitive. However, the mathematical proof is very
novel.

Consider the following Taylor expansion around $c$

\begin{equation*}
f\left( i\right) =f\left( c\right) +f_{i}\left( c\right) \left( i-c\right)
+R, 
\end{equation*}%
where $R$ is the error. We also examine a two-variable function $f\left(
x,y\right) $, however, the extension to a multiple-variable function is
straightforward. As before, we define $h\equiv y+\alpha $, where $\alpha $
is a shift parameter with an initial value equal to zero; so that initially $%
f\left( x+\beta ,y+\alpha \right) \equiv f\left( i,h\right) .$ As before, $f$
is differentiable w.r.t. $h$ and thus $i$ (holding $x$ constant).

The Taylor expansion$\ $is given by 
\begin{equation*}
f\left( i,h\right) =f\left( c_{1},c_{2}\right) +f_{i}\left(
c_{1},c_{2}\right) \left( i-c_{1}\right) +f_{h}\left( c_{1},c_{2}\right)
\left( h-c_{2}\right) +R_{2}. 
\end{equation*}

\textbf{Theorem 1:}

$i)$\textbf{\ }$f\left( i\right) $ is differentiable w.r.t. $i$ (holding $x$
constant).

$ii)$ $f\left( i,h\right) $ is differentiable w.r.t. $i$ and $h$ (holding $x$
and $y$ constant).

\textbf{Proof.}

$i)$ Differentiability with respect to the \textit{shift} \textit{parameter}
(as opposed to a \textit{variable}) stems from the fact that the change in
the shift parameter is a constant, since the function is shifted
horizontally by a constant amount (graphically, this is evidenced by a
horizontal shift of the function). Let $\hat{f}$ be the shifted function and 
$f$ the original function. The impact of the shift at $x$ (holding $x$
constant) is $\hat{f}\left( x\right) -f\left( x\right) $ $=f\left( x+\beta
\right) -f\left( x\right) .$ Let $d\beta =\varphi -0=\varphi $ (since the
initial values are zero), where $\left| \varphi \right| $ is a small
non-zero constant. It is non-zero since the function is shifted. Therefore,
the change in $V$ as a result of the shift holding $x$ constant ($dV/\varphi
)$ can be expressed as this derivative%
\begin{eqnarray*}
f_{i} &\equiv &\frac{df\left( i\right) }{di}\mid _{\triangle x=0}=\underset{%
\triangle i\longrightarrow 0}{\lim }\left[ \frac{f\left( i+\triangle
i\right) -f\left( i\right) }{\triangle i}\mid _{\triangle x=0}\right] \\
&=&\underset{\triangle i\longrightarrow 0}{\lim }\left[ \frac{f\left(
i+\triangle x+\triangle \beta \right) -f\left( i\right) }{\triangle
x+\triangle \beta }\mid _{\triangle x=0}\right] = \\
&&\underset{\triangle i\longrightarrow 0}{\lim }\left[ \frac{f\left(
i+\triangle x+\varphi \right) -f\left( i\right) }{\triangle x+\varphi }\mid
_{\triangle x=0}\right] \\
&=&\frac{f\left( i+\varphi \right) -f\left( i\right) }{\varphi }.
\end{eqnarray*}%
By the continuity of $f$ and the fact that $\varphi \neq 0$, $f_{i}<\infty $
and thus it is differentiable$.\square $

$ii)$ For a two-variable function, the proof is similar. As before, $f\left(
y+\alpha ,x+\beta \right) \equiv f\left( h,i\right) $. Let $\hat{f}$ be the
shifted function and $f$ the original function. The impact of the shift at $%
x $ (holding $x$ (and of course $h)$ constant) is $\hat{f}\left( x,y\right)
-f\left( x,y\right) $ $=f\left( x+\beta ,y\right) -f\left( x,y\right) .$
Thus, consider this derivative

\begin{eqnarray*}
f_{i} &\equiv &\frac{\partial f\left( h,i\right) }{\partial i}\mid
_{\triangle x=0}=\underset{\triangle i\longrightarrow 0}{\lim }\left[ \frac{%
f\left( h+\Delta h,i+\triangle i\right) -f\left( h,i\right) }{\triangle
i+\Delta h}\mid _{\triangle x=\Delta h=0}\right] \\
&=&\underset{\triangle i\longrightarrow 0}{\lim }\left[ \frac{f\left(
h+\Delta h,i+\triangle x+\triangle \beta \right) -f\left( h,i\right) }{%
\triangle x+\triangle \beta +\Delta h}\mid _{\triangle x=\Delta h=0}\right] =
\\
&&\underset{\triangle i\longrightarrow 0}{\lim }\left[ \frac{f\left(
h+\Delta h,i+\triangle x+\varphi \right) -f\left( h,i\right) }{\triangle
x+\varphi +\Delta h}\mid _{\triangle x=\Delta h=0}\right] \\
&=&\frac{f\left( h,i+\varphi \right) -f\left( h,i\right) }{\varphi }.
\end{eqnarray*}%
The same applies to $f_{h}.$ Similarly, the second derivative $f_{ii}=\frac{%
f_{i}\left( h,i+\varphi \right) -f_{i}\left( h,i\right) }{\varphi }$ exists,
since $f_{i}$ is continuous$.\square $

Though we only need local differentiability, this exercise can be repeated
at each value of $x.$

We note that if the function is not differentiable locally (at say $x_{0})$,
our expansion holds locally in the neighborhood of $x_{0}$ (that is, $%
f\left( i\right) =f\left( x_{0}\right) +f_{i}\left( x_{0}\right) \left(
\varphi +x_{0}-x_{0}\right) =f\left( x_{0}\right) +f_{i}\left( x_{0}\right)
\varphi ).$ For the other values/intervals of $x,$ clearly the classical
Taylor expansion holds. Thus, we obtain expansions at both $x_{0}$ and other
values of $x.$ Therefore, the expansion holds generally for $i.$

If the function is no-where differentiable, we repeat the method at each
value of $x.$ Thus, once again, the expansion holds generally.

\textbf{Verification and Examples:}

We provide examples to verify the correctness of the method.

\textit{Example 1}: The absolute value function at zero%
\begin{equation*}
f_{i}=\underset{\triangle i\longrightarrow 0}{\lim }\frac{\left| 0+\varphi
\right| -0}{\varphi }=\frac{\left| \varphi \right| }{\varphi }. 
\end{equation*}
\textit{Example 2: }A no-where differentiable function (the Brownian motion
sample path)

\begin{equation*}
f_{i}=\underset{\triangle i\longrightarrow 0}{\lim }\left[ \frac{\Omega 
\sqrt{dt+\varphi }}{dt+\varphi }\mid _{\triangle t=0}\right] =\frac{\Omega }{%
\sqrt{\varphi }} 
\end{equation*}%
for $\Omega <\infty .$

\bigskip

\section{\protect\bigskip {\protect\large The portfolio model}}

In this paper, we use the standard technical assumptions (except for the
smoothness assumption). We first apply our method to the baseline portfolio
model (see, for example, Cvitanic and Zapatero (20004)). The risk-free asset
price process is given by $S^{0}=e^{rs},$ where $r$ is the constant
risk-free rate of return. The dynamics of the risky asset price are given by

\begin{equation*}
dS_{s}=S_{s}\left( \mu ds+\sigma dW_{s}\right) , 
\end{equation*}%
where $\mu $ and $\sigma $ are the constant rate of return and the
volatility, respectively; $W_{s}$ is a Brownian motion defined on the
probability space $\left( \Omega ,\mathcal{F},\mathcal{F}_{s},P\right) ,$
where $\left\{ \mathcal{F}_{s}\right\} _{t\leq s\leq T}$ is the augmentation
of filtration.

We assume that the wealth process satisfies this equation%
\begin{equation*}
dX_{s}=\left\{ rX_{s}^{\pi }+\left( \mu -r\right) \pi _{s}\right\} ds+\pi
_{s}\sigma dW_{s}, 
\end{equation*}%
or%
\begin{equation*}
X_{T}^{\pi }=x+\int\limits_{t}^{T}\left\{ rX_{s}^{\pi }+\left( \mu -r\right)
\pi _{s}\right\} ds+\int\limits_{t}^{T}\pi _{s}\sigma dW_{s}, 
\end{equation*}%
where $x$ is the initial wealth, $\left\{ \pi _{s},\mathcal{F}_{s}\right\}
_{t\leq s\leq T}$ is the portfolio process, and $E\int\limits_{t}^{T}\pi
_{s}^{2}ds<\infty .$ The trading strategy $\pi _{s}\in \mathcal{A}\left(
x\right) $ is admissible$.$

The investor maximizes the expected utility of the terminal wealth%
\begin{equation*}
V\left( t,x\right) =\underset{\pi }{Sup}E\left[ U\left( X_{T}^{\pi }\right)
\mid \mathcal{F}_{t}\right] , 
\end{equation*}%
where $V\left( .\right) $ is the value function, $U\left( .\right) $ is a
continuous and bounded utility function. It is well known that if $V\left(
t,x\right) $ $\in C^{1,2}\left( \left[ 0,T\right] ,R\right) $, it satisfies
(in the classical sense) the HJB PDE (suppressing the notations)

\begin{equation*}
V_{t}+rxV_{x}+\underset{\pi }{Sup}\left\{ \pi _{t}\left( \mu -r\right) V_{x}+%
\frac{1}{2}\pi _{t}^{2}\sigma ^{2}V_{xx}\right\} =0;V\left( T,x\right)
=U\left( x\right) , 
\end{equation*}%
where the subscripts of $V$ denote partial derivatives. Therefore, the
optimal portfolio is given by

\begin{equation*}
\pi _{t}^{\ast }=-\frac{\left( \mu -r\right) V_{x}}{\sigma ^{2}V_{xx}}. 
\end{equation*}

We define $h$ $\equiv t+\alpha $, $i\equiv x+\beta ,$ and $d\beta $ $%
=\varphi -0=\varphi ,$ where $\alpha $ and $\beta $ are deterministic shift
parameters, each with an initial value equal to zero (see, for example,
Dalal (1990) and Alghalith (2008)); so that initially $V\left( t,x\right)
=V\left( t+\alpha ,x+\beta \right) \equiv V\left( h,i\right) ,$ and $\varphi 
$ is a (small) non-zero constant. Evidently, by construction, $V$ is
continuously differentiable w.r.t. each shift parameter, since any function
can be shifted (graphically the derivative is depicted as a small horizontal
shift of the graph of the function; thus the derivative exists); and hence
it is continuously differentiable w.r.t. $h$ and $i)$, even if it is
non-differentiable w.r.t. $x$ or $t$ (see Appendix 1 for the proof).

In the following proof (and the extensions), we use the standard technical
assumptions (see, for example, Touzi (2002), Touzi (2010) and Strulovici and
Szydlowski (2015) for a discussion), except for the smoothness of value
function.

\bigskip \textbf{Theorem}: The value function $V\left( h,i\right) $
satisfies (in the classical sense) this HJB PDE%
\begin{equation*}
V_{h}+\left( rx+\varphi \right) V_{i}+\underset{\pi _{t}}{Sup}\left\{ \pi
_{t}\left( \mu -r\right) V_{i}+\frac{1}{2}\pi _{t}^{2}\sigma
^{2}V_{ii}\right\} =0,V\left( T,x\right) =U\left( x\right) . 
\end{equation*}%
PROOF\footnote{%
The proof also relies on Appendix 1.}. Define the function $\bar{V}\left(
h,i\right) $ as 
\begin{equation*}
\bar{V}\left( h,i\right) \equiv \bar{V}\left( t,x\right) =E\left[ U\left(
X^{\pi }\left( T\right) \right) /\mathcal{F}_{t}\right] . 
\end{equation*}%
Applying Ito's rule to $\bar{V}\left( h,i\right) ,$ we obtain (suppressing
the notations)

\begin{equation*}
d\bar{V}=\bar{V}_{h}dh+\bar{V}_{i}di+\frac{1}{2}\left( di\right) ^{2}\bar{V}%
_{ii}=\bar{V}_{h}dh+\bar{V}_{i}\left[ dX_{s}+d\beta \right] +\frac{1}{2}%
\left( dX_{s}\right) ^{2}\bar{V}_{ii}= 
\end{equation*}%
\begin{equation*}
\left[ \bar{V}_{h}+\bar{V}_{i}\left( \pi _{s}\left( \mu -r\right)
+rX_{s}^{\pi }+\varphi \right) +\frac{1}{2}\bar{V}_{ii}\pi _{s}^{2}\sigma
^{2}\right] ds+\bar{V}_{i}\pi _{s}\sigma dW_{s}, 
\end{equation*}%
where $h$ and $i$ are defined the same as before (and more generally for
each time $s,$ $h$ $\equiv s+\alpha $, $i\equiv X_{s}+\beta ,$ $d\beta $ $%
=\varphi ).$ Integrating the previous equation yields%
\begin{eqnarray*}
\bar{V}\left( T,X^{\pi }\left( T\right) \right) &=&U\left( X^{\pi }\left(
T\right) \right) =\bar{V}\left( t,x\right) + \\
&&\int\limits_{t}^{T}\left( \bar{V}_{h}+\bar{V}_{i}\left( \pi _{s}\left( \mu
-r\right) +rX_{s}^{\pi }+\varphi \right) +\frac{1}{2}\bar{V}_{ii}\pi
_{s}^{2}\sigma ^{2}\right) ds+ \\
&&\int\limits_{t}^{T}\bar{V}_{i}\pi _{s}\sigma dWs.
\end{eqnarray*}%
Taking expectation expectations on both sides yields%
\begin{eqnarray*}
\bar{V}\left( t,x\right) &=&E\left[ U\left( X^{\pi }\left( T\right) \right) /%
\mathcal{F}_{t}\right] - \\
&&E\left[ \int\limits_{t}^{T}\left( \bar{V}_{h}+\bar{V}_{i}\left( \pi
_{s}\left( \mu -r\right) +rX_{s}^{\pi }+\varphi \right) +\frac{1}{2}\bar{V}%
_{ii}\pi _{s}^{2}\sigma ^{2}\right) ds/\mathcal{F}_{t}\right] .
\end{eqnarray*}%
The above equation implies that for any $\pi _{t}$%
\begin{equation}
\bar{V}_{h}+\left( \pi _{t}\left( \mu -r\right) +rx+\varphi \right) \bar{V}%
_{i}+\frac{1}{2}\pi _{t}^{2}\sigma ^{2}\bar{V}_{ii}=0.  \label{1}
\end{equation}%
Now, by definition

\begin{equation*}
V\left( h,i\right) \equiv V\left( t,x\right) =\text{ }\underset{\pi }{Sup}%
\bar{V}\left( t,x;\pi \right) , 
\end{equation*}%
and thus $\left( \ref{1}\right) $ holds for the optimal portfolio $\pi
_{t}^{\ast }$ 
\begin{equation*}
V_{h}+\left( rx+\varphi \right) V_{i}+\underset{\pi _{t}}{Sup}\left\{ \pi
_{t}\left( \mu -r\right) V_{i}+\frac{1}{2}\pi _{t}^{2}\sigma
^{2}V_{ii}\right\} =0,V\left( T,x\right) =U\left( x\right) .\square 
\end{equation*}%
We also note that integrating over $\left[ 0,x\right] $ and $\left[ 0,t%
\right] $ will yield the original value function $V\left( t,x\right) $ as
the solution. Also, the optimal portfolio is given by%
\begin{equation*}
\pi _{t}^{\ast }=-\frac{\left( \mu -r\right) V_{i}\left( t,x\right) }{\sigma
^{2}V_{ii}\left( t,x\right) }, 
\end{equation*}%
since the derivatives are taken at the initial values $h=t$ and $i=x.$

\section{\protect\large Extensions}

\subsection{\protect\large The portfolio and consumption}

If a part of the wealth can be consumed by the investor (see Hata and sheu
(2012) and Trybola (2015), among others), the wealth process is given by

\begin{equation*}
X_{T}^{\pi ,c}=x+\int\limits_{t}^{T}\left\{ rX_{s}^{\pi ,c}+\left( \mu
-r\right) \pi _{s}-c_{s}\right\} ds+\int\limits_{t}^{T}\pi _{s}\sigma
dW_{s}, 
\end{equation*}%
where $\left\{ c_{s},\mathcal{F}_{s}\right\} _{_{t\leq s\leq T}}$ is the
consumption rate process, with $E\int\limits_{t}^{T}\pi _{s}^{2}ds<\infty $
, $E\int\limits_{t}^{T}c_{s}ds<\infty $ and $c_{s}\geq 0.$ The strategy $%
\left( \pi _{s},c_{s}\right) \in \mathcal{A}$ is admissible. The investor
maximizes the expected utility of the terminal wealth and consumption 
\begin{equation*}
V\left( t,x\right) =\underset{\pi ,c}{Sup}E\left[ U_{1}\left( X_{T}^{\pi
,c}\right) +\int\limits_{t}^{T}U_{2}\left( c_{s}\right) ds\mid \mathcal{F}%
_{t}\right] . 
\end{equation*}%
If it is smooth, the value function satisfies this HJB PDE%
\begin{equation*}
V_{t}+rxV_{x}+ 
\end{equation*}%
\begin{equation*}
\underset{\pi _{t},c_{t}}{Sup}\left\{ \frac{1}{2}\pi _{t}^{2}\sigma
^{2}V_{xx}+\left[ \pi _{t}\left( \mu -r\right) -c_{t}\right]
V_{x}+U_{2}\left( c_{t}\right) \right\} =0, 
\end{equation*}%
\begin{equation*}
V\left( T,x\right) =U\left( x\right) . 
\end{equation*}%
The optimal solutions are%
\begin{equation*}
\pi _{t}^{\ast }=-\frac{\left( \mu -r\right) V_{x}}{\sigma ^{2}V_{xx}}, 
\end{equation*}

\begin{equation*}
U_{2}^{\prime }\left( c_{t}^{\ast }\right) =V_{x}\left( t,x\right) . 
\end{equation*}

Following the previous procedure in Section 2, we can show that the value
function $V\left( h,i\right) $ satisfies (in a classical sense)

\begin{equation*}
V_{h}+\left( rx+\varphi _{1}\right) V_{i}+ 
\end{equation*}%
\begin{equation*}
\underset{\pi _{t},c_{t}}{Sup}\left\{ \frac{1}{2}\pi _{t}^{2}\sigma
^{2}V_{ii}+\left[ \pi _{t}\left( \mu -r\right) -c_{t}\right]
V_{i}+U_{2}\left( g\right) \right\} =0, 
\end{equation*}%
where $g\equiv $ $c_{t}^{\ast }+\gamma ,$ and $\gamma $ is a shift parameter
with an initial value equal to zero (defined the same as before) and $%
d\gamma =\varphi _{1}$ is a non-zero constant; also, $i$ and $h$ are defined
the same as before. Thus, the optimal solutions are%
\begin{equation*}
\pi _{t}^{\ast }=-\frac{\left( \mu -r\right) V_{i}\left( t,x\right) }{\sigma
^{2}V_{ii}\left( t,x\right) }, 
\end{equation*}

\begin{equation*}
U_{2}^{\prime }\left( c_{t}^{\ast }\right) =V_{i}\left( t,x\right) . 
\end{equation*}

\subsection{\protect\large The portfolio with a (stochastic) economic factor}

The stochastic factor model assumes that the rate of return and volatility
are functions of a stochastic (economic) factor (see, for example, Alghalith
(2009) and Trybola (2015)). This implies a two-dimensional standard Brownian
motion $\left\{ \left( W_{s}^{1},W_{s}^{2}\right) ,\mathcal{F}_{s}\right\}
_{t\leq s\leq T}$ defined on the probability space $\left( \Omega ,\mathcal{F%
},\mathcal{F}_{s},P\right) $. The risk-free asset price process is $%
S^{0}=e^{rs},$ where $r$ is the rate of return and $Y_{s}$ is the economic
factor.

The risky asset price process is given by

\begin{equation*}
dS_{s}=S_{s}\left\{ \mu \left( Y_{s}\right) ds+\sigma \left( Y_{s}\right)
dW_{s}^{1}\right\} , 
\end{equation*}%
where $\mu \left( Y_{s}\right) $ and $\sigma \left( Y_{s}\right) \in
C_{b}^{2}\left( R\right) $ are the rate of return and the volatility,
respectively. The economic factor process satisfies 
\begin{equation*}
dY_{s}=b\left( Y_{s}\right) ds+\rho dW_{s}^{1}+\sqrt{1-\rho ^{2}}%
dW_{s}^{\left( 2\right) },Y_{t}\equiv y, 
\end{equation*}%
where $\left| \rho \right| <1$ is the correlation factor between the two
Brownian motions and $b\left( Y_{s}\right) \in C^{1}\left( R\right) $.

The wealth process satisfies

\begin{equation*}
X_{T}^{\pi }=x+\int\limits_{t}^{T}\left\{ rX_{s}^{\pi }+\left[ \mu \left(
Y_{s}\right) -r\right] \pi _{s}\right\} ds+\int\limits_{t}^{T}\pi _{s}\sigma
\left( Y_{s}\right) dW_{s}^{1}. 
\end{equation*}%
The investor maximizes the expected utility of the terminal wealth 
\begin{equation*}
V\left( t,x,y\right) =\underset{\pi }{Sup}E\left[ U\left( X^{\pi }\right)
\mid \mathcal{F}_{t}\right] . 
\end{equation*}%
If it is smooth, the value function satisfies this Hamilton-Jacobi-Bellman
PDE

\begin{equation*}
V_{t}+rxV_{x}+b\left( y\right) V_{y}+\frac{1}{2}V_{yy}+ 
\end{equation*}%
\begin{equation*}
\underset{\pi _{t}}{Sup}\left\{ \frac{1}{2}\pi _{t}^{2}\sigma ^{2}\left(
y\right) V_{xx}+\left[ \pi _{t}\left( \mu \left( y\right) -r\right) \right]
V_{x}+\rho \sigma \left( y\right) \pi _{t}V_{xy}\right\} =0, 
\end{equation*}%
\begin{equation*}
V\left( T,x,y\right) =U\left( x\right) . 
\end{equation*}%
Hence, the optimal portfolio is

\begin{equation*}
\pi _{t}^{\ast }=-\frac{\left[ \mu \left( y\right) -r\right] V_{x}}{\sigma
^{2}\left( y\right) V_{xx}}-\frac{\rho V_{xy}}{\sigma \left( y\right) V_{xx}}%
. 
\end{equation*}

Using the previous procedure, we can show that the value function $V\left(
h,i,j\right) $ satisfies (in a classical sense) this HJB PDE%
\begin{equation*}
V_{h}+\left( rx+\varphi _{2}\right) V_{i}+\left( b\left( y\right) +\iota
\right) V_{j}+\frac{1}{2}V_{jj}+ 
\end{equation*}%
\begin{equation*}
\underset{\pi _{t}}{Sup}\left\{ \frac{1}{2}\pi _{t}^{2}\sigma ^{2}\left(
y\right) V_{ii}+\left[ \pi _{t}\left( \mu \left( y\right) -r\right) \right]
V_{i}+\rho \sigma \left( y\right) \pi _{t}V_{ij}\right\} =0, 
\end{equation*}%
\begin{equation*}
V\left( T,x,y\right) =U\left( x\right) , 
\end{equation*}%
where $j\equiv y+\zeta ,$ and $\zeta $ is a shift parameter with an initial
value equal to zero (defined the same as before) and $d\zeta =\varphi _{2}$
is a non-zero constant; $h$ and $i$ are defined the same as before.
Therefore, the optimal solution is%
\begin{equation*}
\pi _{t}^{\ast }=-\frac{\left[ \mu \left( y\right) -r\right] V_{i}\left(
t,x,y\right) }{\sigma ^{2}\left( y\right) V_{ii}\left( t,x,y\right) }-\frac{%
\rho V_{ij}\left( t,x,y\right) }{\sigma \left( y\right) V_{ii}\left(
t,x,y\right) }. 
\end{equation*}

\textbf{Appendix 1. }Proof of the differentiability

Differentiability with respect to the \textit{shift} \textit{parameter} (as
opposed to a \textit{variable}) stems from the fact that the change in the
shift parameter is a constant (graphically, this is evidenced by a
horizontal shift of the function). As before, $V\left( t+\alpha ,x+\beta
\right) \equiv V\left( h,i\right) ,$ and let $d\alpha =\epsilon -0=\epsilon
,d\beta =\varphi -0=\varphi $ (since the initial values are zero), where $%
\epsilon $ and $\varphi $ are small non-zero constants. Consider this
derivative

\begin{eqnarray*}
\frac{\partial V\left( h,i\right) }{\partial i} &\mid &_{\triangle x=0}=%
\underset{\triangle i\longrightarrow 0}{\lim }\frac{V\left( h,i+\triangle
i\right) -V\left( h,i\right) }{\triangle i}\mid _{\triangle x=0} \\
&=&\underset{\triangle i\longrightarrow 0}{\lim }\frac{V\left( h,i+\triangle
x+\triangle \beta \right) -V\left( h,i\right) }{\triangle x+\triangle \beta }%
\mid _{\triangle x=0}= \\
&&\underset{\triangle i\longrightarrow 0}{\lim }\frac{V\left( h,i+\triangle
x+\varphi \right) -V\left( h,i\right) }{\triangle x+\varphi }\mid
_{\triangle x=0} \\
&=&\frac{V\left( h,i+\varphi \right) -V\left( h,i\right) }{\varphi }.
\end{eqnarray*}%
By the continuity and boundedness of $V$ and the fact that $\varphi \neq 0$,
the derivative exists. Since $x$ and $\beta $ are independent ($\frac{dx}{%
d\beta }$ $=0),$ $\frac{\partial V\left( h,i\right) }{\partial i}\mid
_{\triangle x=0}=\frac{\partial V\left( h,i\right) }{\partial i}\equiv
V_{i}. $ Similarly, $\frac{\partial V\left( h,i\right) }{\partial h}\mid
_{\triangle t=0}=\frac{\partial V\left( h,i\right) }{\partial h}\equiv
V_{h}, $ since $\frac{dt}{d\alpha }$ $=0.$ Similarly, the second derivative $%
V_{ii}=\frac{V_{i}\left( h,i+\varphi \right) -V_{i}\left( h,i\right) }{%
\varphi }$ exists, since $V_{i}$ exists and $\varphi \neq 0.\square $Under
the assumption of a non-constant marginal utility, $V_{ii}\neq 0.$

\end{document}